\documentclass[12pt]{article}
\usepackage{latexsym,amssymb,amsmath,amscd,amscd,amsthm,amsxtra}
\usepackage[dvips]{graphicx}
\newtheorem{Theorem}{Theorem}[section]

\def\ba{\begin{array}}
\def\ea{\end{array}}
\def\be{\begin{equation}\begin{array}{l}}
\def\ee{\end{array}\end{equation}}
\def\c{\cite}
\def\a{\alpha}

\def\pa{\partial}

\begin{document}
\begin{flushright}
AS-ITP-2001-010\\
April 17, 2001   \\
\end{flushright}
\vskip 2cm
\noindent
\begin{center}
 {\large\bf   On  Symplectic, 
Multisymplectic Structures-Preserving \\
\vskip 3mm
in  Simple
 Finite Element Method}

\end{center}
\vspace*{1cm}

\def\thefootnote{\fnsymbol{footnote}}
\begin{center}{\sc Han-Ying GUO${}^1$}\footnote{email: hyguo@itp.ac.cn},
{\sc Xiao-mei JI ${}^{1,2}$}\footnote{email: jixm@iu-math.math.indiana.edu }
{\sc Yu-Qi LI${}^{1}$}\footnote{email: qylee@itp.ac.cn}
 and {\sc Ke WU${}^1$}\footnote{email: wuke@itp.ac.cn} 
\end{center}
\begin{center}
{\it ${}^1$ Institute of Theoretical Physics, Academia Sinica, P.O.Box 2735,\\ Beijing 
100080, China \par
${}^2$ Department of Mathematics, Indiana University\\ Bloomington, IN 47405, U.S.A.}
\end{center}

\vskip 2mm
\vfill

\noindent
\centerline {\sc Abstract}
\vskip 3mm

$\quad $ By the simple finite element method, we  study the
symplectic, multisymplectic structures and relevant preserving properties
in some semi-linear
elliptic boundary value
problem in one-dimensional and two-dimensional spaces respectively. We
find that with uniform mesh, the numerical schemes derived
from finite element method can keep a preserved symplectic structure
in one-dimensional case and a preserved multisymplectic structure in
two-dimentional case respectively.
These results are in fact the intrinsic reason
that the numerical experiments indicate
that such finite element schemes are accurate in practice.

\vskip .8cm
 
\vskip .8cm

\noindent

\noindent
{\sl Keywords:} symplectic structure, finite element method, numerical scheme
\vskip .4cm
\newpage

\section{Introduction}

 $\quad $   It is well-known that the advantage of the finite element
method is to adapt the flexible and complicated unstructured grids, whose
mathematical theory is more profound and complete \cite{fini},
and the systems to deal with by the method are Lagrangian ones in general.
 On the other hand, the symplectic algorithm \cite{symp} and multisymplectic
 algorithm \cite{msymp} are also very
powerful and successful for the finite and infinite dimensional Hamiltonian
systems respectively in comparison with other non-symplectic and/or
non-multi symplectic schemes.
Therefore, there is a very natural but important and intriguing problem.
Namely, what is the relation between these two extremely important aspects
in scientific computing. More concretely, whether there exist some
preserved symplectic and/or multisymplectic structures in finite element
schemes. As far as we know, it is still an open problem.

   In this paper, we study and definitely answer this problem by exploring 
certain concrete samples. We consider the boundary value problem of
semi-linear elliptic equation with ordinary finite element schemes. In this
method, we use linear elements for the spatial discretization. With uniform
mesh, we have found that there exist certain preserved symplectic
structure in one-dimensional space and certain multisymplectic
structure that is also preserved in the discretized sense of the
finite element method in two-dimensional space.

In the structure-preserving point of view, these results in fact are 
intrinsic reason why the numerical experiments 
state that such finite element algorithms are accurate in practice.

In what follows, we first consider the symplectic and multisymplectic 
geometry for the semi-linear elliptic equation in one-dimensional and
two-dimensional spaces for both continuous and difference discrete cases 
in section 2. By introducing the Euler-Lagrange (EL) 1-forms, i.e. the
null EL 1-form is given as the equation producting by a certain relevant 
1-form which is cohomologically equivalent to the coboundary EL 1-forms, 
and EL condition, i.e. the closed EL 1-forms as well as
their difference discrete versions \c{glw1} \c{glw2}, we show that the 
symplectic
structure in one-dimensional case and the multisymplectic structure in 
two-dimensional case and their difference discrete versions are preserved
in the relevant configuration spaces in general rather than in the solution 
spaces of the equations only. In section 3, we briefly introduce relevant
issues by 
the finite element scheme in one-dimensional and two-dimensional
spaces for the boundary problem of the semi-linear elliptic equation. In
section 4, we study the discrete versions of the symplectic and
multisymplectic 
structures in the one-dimensional and two-dimensional finite element 
method for the semi-linear elliptic equation given in the section 3. 
We find explicitly the 
discrete versions of the symplectic and multisymplectic 
structures  and their preserving properties respectively. Finally, we end
with some remarks.

\section{Symplectic geometry for semi-linear elliptic equation in
1-D and 2-D spaces}

 $\quad $   In this paper, we consider the boundary value problem of the
 semi-linear elliptic equation in one-dimensional and two-dimensional spaces:
\begin{equation}
\triangle u=f(u) \quad in \quad \Omega,
\quad u|_{\partial{\Omega}}=0 \quad on \quad \partial\Omega.
\end{equation}
$\Omega $ is a bounded domain in ${I\!\!R}^n$, $n=1,2$ and $f(u)$
is nonlinear and sufficiently smooth enough.

The weak formulation of the boundary value problem of the equation is:
to find $u:\Omega\rightarrow H^1_0(\Omega)$ such that
\begin{equation}
\label{30}
\int _{\Omega}\nabla u\cdot\nabla v dx=
-\int _{\Omega}f\cdot v dx \hskip
0.5cm \forall v\in H^1_0(\Omega).
\end{equation}
Let
$$
a(u,v)=\int _{\Omega}\nabla u\cdot\nabla v dx,\quad
(f,v)=-\int _{\Omega}f\cdot v dx,
$$
then (\ref{30}) becomes,
\begin{eqnarray}
a(u,v)= (f,v).
\end{eqnarray}
It is important to note that in the one-dimensional case the equation is in
fact an ODE with Lagrangian on
${I\!\!R}^1$. In the two-dimensional case the equation is also the one with
Lagrangian. Therefore, in the continuous cases, they should be symplectic
structure-preserving and multisymplectic structure-preserving respectively
\c{symp} \c{msymp} \c{glw1}.
In this section, we briefly recall these facts with the help
of the relevant Euler-Lagrange (EL) cohomology\c{glw1}.
For the details of the symplectic and multisymplectic structures in the
Lagrangian formalism, it can be found in \c{mars} \c{qin} \c{glw1}. The
discrete versions of this
kind of symplectic and multisymplectic structures are also derived from
discrete variational principle
\c{vese} \c{mars} \c{qin} \c{glw1}.

\subsection{Symplectic structure in 1-D space case} 

 $\quad $   In one-dimensional space, the semi-linear equation
 becomes
\begin{equation}
 u_{xx}=f(u) \quad in \quad \Omega,
\quad u|_{\partial{\Omega}}=0 \quad on \quad \partial\Omega.
\end{equation}
where $u_x$ denotes the derivative of $u$ with respect to $x$,
which is the coordinate of $\Omega$ and $\Omega $ a segment bounded domain in
${I\!\!R}^1$.

Let us first release both $u$ and $u_x$ from the solution space of the
equation to
the function space on $\Omega$.
In order to do so, we introduce the  
EL 1-forms in the function space \c{glw1}
\begin{equation}
E(u, u_x):=\{u_{xx}-f(u)\}du,
\end{equation}
and the EL condition \c{glw1}, i.e. the EL 1-form is closed
\begin{equation}
dE(u, u_x)=0,
\end{equation}
leads to the symplectic structure and its preserving law.

It is easy to see that the null EL 1-form is equivalent to the equation. 
And it is a special case of the coboundary EL 1-forms given by,
\begin{equation}
E(u, u_x)=d\a (u, u_x),
\end{equation}
where $\a (u, u_x )$ is an arbitrary function of $(u, u_x)$ in the function 
space, automatically satisfying the  EL condition. 
It is cohomologically trivial and equivalent to the null 1-form. Therefore,
 $u$ and $u_x$ in the EL 1-forms $E(u, u_x)$  are 
{\it Not} in the solution 
space of the equation in general. 
  
We may now introduce a new variable
$$
v=u_x
$$
in the function space and the EL condition becomes
\begin{equation}
v=u_x, \quad d\{\{v_x-f(u)\}du\}=0, 
\end{equation}
Now it is easy to prove that the EL condition in the
function space leads directly to the following equation:
\begin{equation}
\frac {d} {dx} \{dv \wedge du\}=0.
\end{equation}
This means that there exists an intrinsic symplectic 2-form
\begin{equation}
\omega=dv \wedge du,
\end{equation}
and it is divergence free in the sense 
\begin{equation}
\frac {d} {dx} \omega =0.
\end{equation}
In other words, the one-dimensional equation under consideration is
symplectic structure-preserving.

We may also eliminate the variable $v$ to get symplectic
2-form in the Lagrangian formalism
 for the one-dimensional semi-linear elliptic equation as follows:
\begin{equation}
\omega=du_x \wedge du,
\end{equation}
which is also preserved in the sense of divergence free
\begin{equation}
\frac {d} {dx} \omega =0.
\end{equation}

In order to discretize the equation, we may take the difference with equal
spatial step $h$ for the independent variable $x$, 
$$
x^{(n+1)}=x^{(n)}+h, \quad n=0,\cdots,N.$$
Then we may get the numerical integrator for the equation. For example,
\begin{equation}
u^{(n+1)}-u^{(n)}=hv^{(n)}, \quad v^{(n+1)}-v^{(n)}=hf(u^{(n+1)}).
\end{equation}
It is symplectic structure-preserving scheme in the
following sense
\begin{equation}
dv^{(n+1)}\wedge du^{(n+1)}=dv^{(n)}\wedge du^{(n)}.
\end{equation}
We will explain this issue later.

For the discrete case in difference, the discrete 
Euler-Lagrange (DEL) equation or the
numerical scheme  can be written as
\begin{equation}
u^{(n+2)}-2u^{(n+1)}+u^{(n)}=h^2 f(u^{(n+1)}).
\end{equation}
It can be proved that it is discretely symplectic 
structure-preserving \c{glw1}. The major points are as follows.

Introducing the DEL 1-form
\begin{equation}
E_D^{(n+1)}:=\{u^{(n+2)}-2u^{(n+1)}+u^{(n)}-h^2 f(u^{(n+1)}) \}du^{(n+1)},
\end{equation}
the null DEL 1-form gives the DEL equation and also is a special case of
the coboundary DEL 1-forms
\begin{equation}
E_D^{(n+1)}=d\a_D^{(n+1)}(u^{(n+1)}),
\end{equation}
where $\a_D^{(n+1)}(u^{(n+1)})$ is an arbitrary function of $ u^{(n+1)}$.
Let us consider  the DEL condition, i.e. the closed DEL 1-forms
\begin{equation}
dE_D^{(n+1)}=0.
\end{equation}
It is straightforward to prove that from the DEL 
condition, 
it follows the symplectic preserving property:
\begin{equation}
du_x^{(n+1)}\wedge du^{(n+1)}=du_x^{(n)}\wedge du^{(n)},
\end{equation}
where $u_x^{(n)}$ denotes the difference of $u$ at $x^{(n)}$ and
the discrete symplectic 2-form is given by\c{vese} \c{mars}
\c{qin} \c{glw1} \c{glw2},
\begin{equation}
\omega_D^{(n)}=du_x^{(n)}\wedge du^{(n)}.
\end{equation}

\subsection{Multisymplectic structure in 2-D space case} 

 $\quad $   In the two-dimensional case, the semi-linear equation becomes
\begin{equation}
 u_{x_1x_1}+u_{x_2x_2}=f(u) \quad in \quad \Omega,
\quad u|_{\partial{\Omega}}=0 \quad on \quad \partial\Omega.
\end{equation}
where $u_{x_i}$, $i=1,2$, denote the partial derivative of
$u$ w.r.t. $x_i$, $x_i$ the coordinates of $\Omega$ and $\Omega$ a
bounded domain in ${I\!\!R}^2$.

Similar to the case of one-dimensional space, we can also introduce the 
EL 1-form,
\begin{equation}
E(u, u_{x_i}):=\{ u_{x_1x_1}+u_{x_2x_2}-f(u)\}du,
\end{equation}
such that the null EL 1-form is equivalent to the equation. And the
coboundary EL
1-forms, i.e.
$$
E(u, u_{x_i})=d\a (u, u_{x_i}),
$$
where $\a (u, u_{x_i})$ is an arbitrary function of $(u, u_{x_i})$, 
 are cohomologically equivalent to the null EL 1-form and  they are
cohomologically trivial.

Let us consider the following EL condition, i.e. the closed EL 1-forms
\begin{equation}
dE(u, u_{x_i})=0.
\end{equation}
It is  easy to see that 
$u(x_i)$'s in the cohomological class are not the solution
of the equation in general. 

We may introduce two new variables in the
function space on $\Omega$
\begin{equation}
v=u_{x_1}, \quad  w=u_{x_2}.
\end{equation}
Then the EL condition becomes
\begin{equation}
v=u_{x_1},\quad  w=u_{x_2}, \quad 
d\{\{\frac{\pa}{\pa x_1} v +\frac{\pa}{\pa x_2}w - f(u)\}du\}=0,
\end{equation}
where 
 $(u, v, w)$ are in the function
space  on $\Omega$ rather than in the solution space of the equation only. 
By making use of the nilpotency of $d$, i.e.
 $d^2=0,$
 it follows from the EL condition that
\begin{equation}
\frac {\pa} {\pa x_1} \{ dv \wedge du\}
+\frac {\pa} {\pa x_2} \{ dw \wedge du\}=0.
\end{equation}
This means that there are two intrinsic symplectic 2-forms
\begin{equation}
\omega=dv \wedge du,\quad \quad 
\tau=dw \wedge du.
\end{equation}
 and they are preserved in the sense of divergence free 
\begin{equation}
\frac {\pa} {\pa x_1} \omega +\frac {\pa} {\pa x_2} \tau =0.
\end{equation}
Therefore, the two-dimensional equation under consideration has what is called 
multisymplectic structure-preserving law that holds
 not only in the solution space but also
in the function space under condition (24) as well. 

For the difference discretized version, one of the discrete versions for the
 DEL equations or the 
numerical schemes can be given by,
\begin{equation}
\frac{u_{(i+1,j)}-2u_{(i,j)}+u_{(i-1,j)}}{{\Delta x_1}^2}+
\frac{u_{(i,j+1)}-2u_{(i,j)}+u_{(i,j-1)}}{{\Delta x_2}^2}
 =f(u_{(i,j)}),
\end{equation}
where $\Delta x_i$, $i=1,2$ are difference step-length in $x_i$.
 
We can also release $u_{(i,j)}$ 's from the solution space to the
function space
by introducing the DEL 1-forms 
\begin{equation}
E_D:=\{\frac{u_{(i+1,j)}-2u_{(i,j)}+u_{(i-1,j)}}{{\Delta x_1}^2}+
\frac{u_{(i,j+1)}-2u_{(i,j)}+u_{(i,j-1)}}{{\Delta x_2}^2}
 -f(u_{(i,j)})\}du_{(i,j)},
\end{equation}
and the null DEL 1-form is equivalent to the 
DEL equation  as in the continuous 
case \c{glw1}. For the coboundary DEL 1-forms which are cohomologically 
trivial and equivalent to the null DEL 1-form. Then for the DEL 1-forms,
specially for the closed DEL 1-forms, i.e. they
satisfy the DEL condition
\begin{equation}
dE_D=0,
\end{equation}
the $u_{(i,j)}$'s are in the function space in general rather than in
the solution space only. 
Then it is straightforward from the DEL condition to get
the discrete divergence free equations, i.e.
the discrete multisymplectic structure-preserving property as follows:
\begin{equation}
\frac{1}{{\Delta x_1}^2}D_{x_1}(du_{x_1(i,j)}\wedge du_{(i,j)})+
\frac{1}{{\Delta x_2}^2}D_{x_2}(du_{x_2(i,j)}\wedge du_{(i,j)})=0,
\end{equation}
where $u_{x_i}$ is the difference of $u$ along $x_i$.  These differences can
be expressed as 
$$
D_{x_1}=E_{x_1}-1, \quad D_{x_2}=E_{x_1}-1,
$$
and
$$
E_{x_1}u_{(i,j)}=u_{(i+1,j)}, \quad E_{x_2}u_{(i,j)}=u_{(i,j+1)}.
$$

\section{Finite element method in 1-D and 2-D spaces}

 $\quad $   
In this section, we consider certain
spacial discretization and derive the relevant finite element
schemes for the equation (1) in one-dimensional and two-dimensional cases. The problem on existence and
uniqueness of the (weak) solution of schemes  derived from finite element method for
the semi-linear elliptic  boundary problem will be given
in  \cite{ji}.
In the next section we  will
show that the derived 
schemes in one-dimensional and two-dimensional cases
are  discretely symplectic and multisymplectic structure-preserving
respectively.

\subsection{Finite element scheme in one-dimensional space case} 

   $\quad $    Let $\Omega$ be a segment, $x_i$  the node,
    $\varphi _i$ shape function such that $\varphi _i(x_j)=\delta _{ij}$,
    ${\cal T}$  the set of elements neighboring  for given $x_i$,
$\Omega _i=\cup _{T\in {\cal T}}T$, and $I_T$  the index set of
the nodes of $T$ besides $x_i$. As usual, $u_i=u(x_i)$.
Let 
$$
u_h=\sum_{i=1}^{N_1}u_i\varphi_i,\quad v_h=\sum_{i=1}^{N_1}v_i\varphi_i,
$$
 so
$$
a(u_h,v_h)=a(\sum_{j=1}^{N_1}u_j\varphi_j,\sum_{i=1}^{N_1}v_i\varphi_i),\quad
(f,v_h)=(f,\sum_{i=1}^{N_1}v_i\varphi_i).
$$
By the definition of weak solution,
$a(u_h,v_h)=(f,v_h)$,  $\forall v_i$, we obtain
$$a(\sum_{j=1}^{N_1}u_j\varphi_j,\varphi_i)=(f,\varphi_i),\quad i=1,2,...,N_1.$$
we can get the finite element scheme as follows
\begin{equation}
\frac{u_{i+1}-2u_i+u_{i-1}}{h}=-(f(\sum_{i-1}^{i+1}u_k\varphi_k), \varphi_i).
\end{equation}

\subsection{Finite element scheme in two-dimensional space case}

 $\quad $   Let $\Omega$ be a bounded polygonal domain in ${I\!\!R}^2$,
 $x_{i,j}$ 
 the nodes, and
$\varphi_{i,j}$ shape function such that
$\varphi_{i,j}(x_{k,l})=\delta_{i,k}\delta_{j,l}$.
For given
$x_{i,j}$, let ${\cal T}$ be the set of elements neighboring $x_{i,j}$,
$\Omega _{i,j}=\cup _{T\in {\cal T}}T$, 
and 
$u(x_{i,j})=u_{i,j}$. Similar to one-dimensional
case, we can get the finite element scheme:
\begin{equation}
a(\sum_{i,j=1}^{N_1}u_{i,j}\varphi_{i,j},\varphi_{i,j})
=-(f,\varphi_{i,j}),\quad  i,j=1,2,...,N_1.
\end{equation}
 For $n=2$ we assume $\Omega$ is a square
domain, and the mesh
is uniform, that is, the plane ${I\!\!R}^2$ is divided into squares
$\{x;i_1h\leq x_1\leq (i_1+1)h, i_2h\leq x_2\leq (i_2+1)h\},
i_1,i_2=0,\pm 1,\pm 2
,\cdots$, and each square is further divided into two triangles by a
straight line $x_2=x_1+ih$, $i$ integer.
Take the node $x_{i,j}$ and $\Omega _{i,j}$ shown as figure 1. The elements
are divided  into two categories: the first  category is shown as figure 2
and the second is shown as figure 3.
  For each element in the first category, the element stiff matrix is:
 $$\displaystyle (a_{ij})=\begin{pmatrix}
&\frac12&-\frac12&0  \\
&-\frac12&1&-\frac12  \\
&0&-\frac12&\frac12    \\
\end{pmatrix},$$
For each element in the second category, the element stiff matrix
 is 
 $$\displaystyle (a_{ij})=\begin{pmatrix}
&\frac12&0&-\frac12  \\
&0&\frac12&-\frac12   \\
&-\frac12&-\frac12&1    \\
\end{pmatrix}.$$
We can get the finite element scheme as follows
\begin{eqnarray}
\label{fin}
        u_{i,j+1}+u_{i+1,j}+u_{i,j-1}+u_{i-1,j}-4u_{i,j}\nonumber\\
=\int_{\Omega_{i,j}}
f(\sum u_{k,l}\varphi_{k,l})\varphi_{i,j}dx.
\end{eqnarray}

\begin{figure}
\begin{center}
\begin{minipage}{5cm}
\includegraphics[width=5cm]
{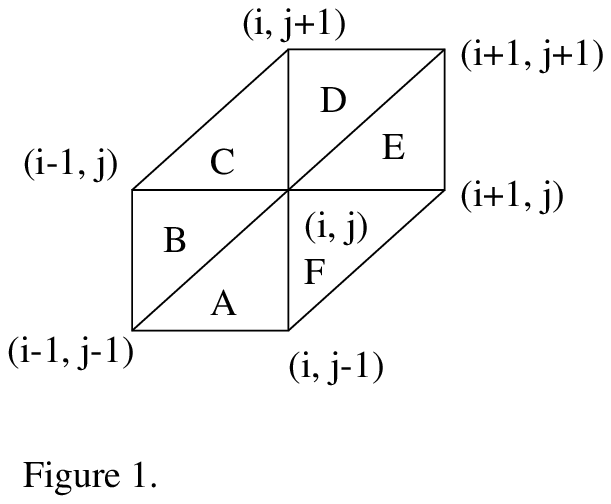}
\caption{}

\end{minipage}
\hskip3mm
\begin{minipage}{3.5cm}
\includegraphics[width=3.5cm]
{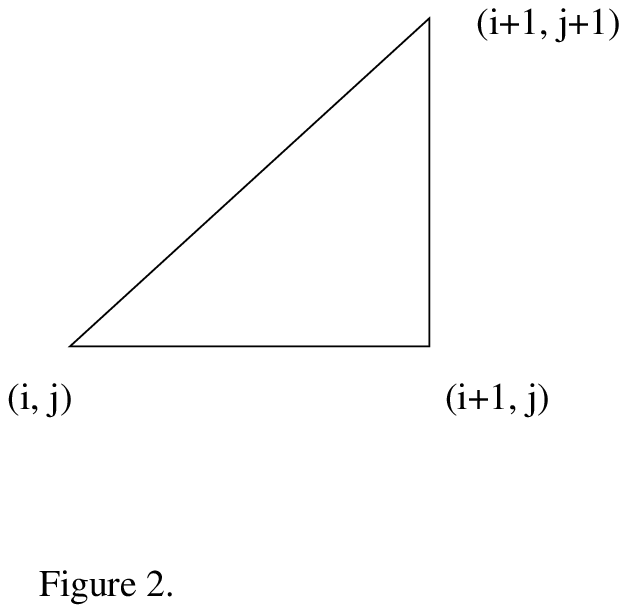}
\caption{}

\end{minipage}

\begin{minipage}{3.5cm}
\includegraphics[width=3.5cm]
{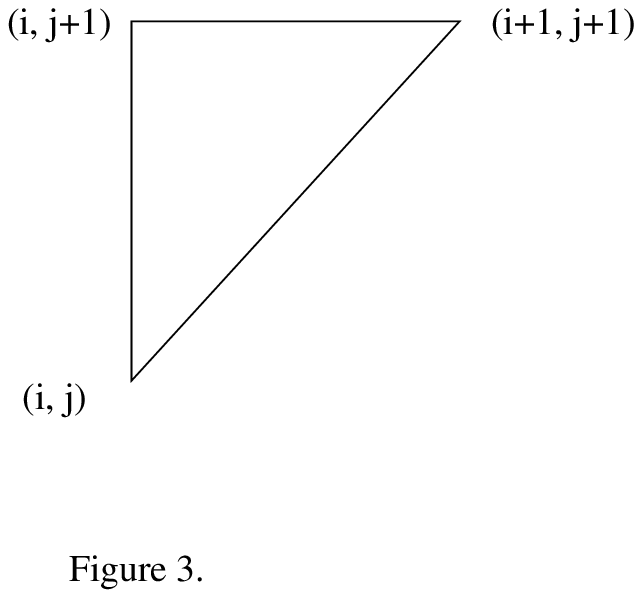}
\caption{}

\end{minipage}
        
\end{center}
\end{figure}

\section{The symplectic and multisymplectic structures in the
1-D and 2-D finite element methods}

\subsection{The symplectic structure in 1-D space}

 $\quad $   In the case of 1-dimensional space, the equation (1) becomes an
 ODE as follows
 \begin{eqnarray}
        u_{xx}=f(u)  \quad u|_{\partial{\Omega}}=0,
\end{eqnarray}
where $x\in \Omega$, $\Omega$ is a segment of ${I\!\!R}^1$ with regular
discretization of  equal spatial step $h$.
                    
From the finite element method, we get the scheme in last section
\begin{eqnarray}
\frac{u_{i+1}-2u_i+u_{i-1}}{h}=
\int_{i-1}^{i+1}f(\sum_{i-1}^{i+1}u_k\varphi_k)\varphi_idx.
\end{eqnarray}
The right side can be rewritten as
\begin{eqnarray}
I_i:=\int_{i-1}^{i}f(u_{i-1}\varphi_{i-1}+u_{i}\varphi_{i})\varphi_idx+
\int_{i}^{i+1}f(u_{i}\varphi_{i}+u_{i+1}\varphi_{i+1})\varphi_idx.
\end{eqnarray}

It should be noted as before that the variables $u_i$'s in the scheme 
can be released from the solution
space to the function space by means of relevant DEL cohomologically
equivalent relation.
Therefore, as long as working with the DEL cohomology class associated with
 the DEL
equation, the $u_{k}$'s can be regarded as in
the function space in general rather than in the solution space only. 

Introducing the DEL 1-forms
\begin{eqnarray}
E_{D i} :=\{{u_{i+1}-2u_i+u_{i-1}}-hI_i\}du_i,
\end{eqnarray}
such that the null DEL 1-form gives rise to the equation in the finite
element
method. Considering the coboundary DEL 1-form
\begin{eqnarray}
E_{D i} =d\a_{D i},
\end{eqnarray}
where $ \a_{D i}$ an arbitrary function  on the function space, the 
null DEL 1-form is a special case of them. The DEL condition reads
\begin{eqnarray}
dE_{D i} =0. 
\end{eqnarray}
Namely, the DEL 1-forms are closed. 
Then it is easy to see that the variables $u_i$'s in the DEL 1-forms are
in the function space in general and it is straightforward to see that
from the DEL condition it follows
\begin{eqnarray}
&&du_{i+1}\wedge du_i+du_{i-1}\wedge du_i
\nonumber\\
&=&h(\int_{i-1}^{i}f^{\prime}(u_{i-1}\varphi_{i-1}+u_{i}\varphi_{i})
\varphi_{i-1}\varphi_idx)du_{i-1}\wedge du_i\nonumber\\
&\quad +&h(\int_{i}^{i+1}f^{\prime}(u_{i}\varphi_{i}+u_{i+1}\varphi_{i+1})
\varphi_{i+1}\varphi_idx)du_{i+1}\wedge du_i ,
        \end{eqnarray}
i.e.,
\begin{eqnarray}
&&(1-h\int_{i}^{i+1}f^{\prime}(u_{i}\varphi_{i}+u_{i+1}\varphi_{i+1})
\varphi_{i+1}\varphi_idx)du_{i+1}\wedge du_i\nonumber\\
&=&(1-h\int_{i-1}^{i}f^{\prime}(u_{i-1}\varphi_{i-1}+u_{i}\varphi_{i})
\varphi_{i-1}\varphi_idx)du_{i}\wedge du_{i-1} .
        \end{eqnarray}

Let us define a 2-form as
\begin{eqnarray}
\omega^{(i+1)}
=(1-h\int_{i}^{i+1}f^{\prime}(u_{i}\varphi_{i}+u_{i+1}\varphi_{i+1})
\varphi_{i+1}\varphi_idx)du_{i+1}\wedge du_i.
\end{eqnarray}
It can be checked that this 2-form is closed with respect to  $d$
on the function space and its coefficients are non-degenerate,
so that it is a symplectic structure for the scheme derived
from finite element method in the case of one-dimensional space and
it is preserved
in the sense
\begin{eqnarray}
\omega^{(i+1)}=\omega^{(i)}.
 \end{eqnarray}

 \subsection{Multisymplectic structure in 2-D space case}

 $\quad $   Now we discuss the following semi-linear equation
\begin{eqnarray}
        u_{x_1x_1}+u_{x_2x_2}=f(u),
\end{eqnarray}
where $x_1, x_2 $ are coordinates in ${I\!\!R}^2$. We first
discrete ${I\!\!R}^2$ with regular lattice
${I\!\!L}^2={I\!\!L}^1 \otimes {I\!\!L}^1$,
and ${I\!\!L}^1$ is  of equal spatial step $h$.

The discrete scheme for this equation form finite element method is
given by (36) as follows
\begin{eqnarray}
\label{fin0}
        u_{i,j+1}+u_{i+1,j}+u_{i,j-1}+u_{i-1,j}-4u_{i,j}\nonumber\\
=\int_{\Omega_{i,j}}
f(\sum u_{k,l}\varphi_{k,l})\varphi_{i,j}dx,  \nonumber
\end{eqnarray}
where the shape function $\varphi_{i,j}$ is linear continuous function
defined as
$$
\varphi_{i,j}(x_{k,l})=\delta_{i.k}\delta_{j,l}.
$$
The right side of (36) can be denoted by
\begin{eqnarray}
I_{\Omega_{i,j}}&:=&\int_{\Omega_{i,j}} f(\sum u_{k,l}\varphi_{k,l})
\varphi_{i,j}dx\nonumber\\
&=&(\int_A+\int_B+\int_C+\int_D+\int_E+\int_F)f(\sum u_{k,l}\varphi_{k,l})
\varphi_{i,j}dx,
\end{eqnarray}
where
\begin{eqnarray}
I_{\Omega_{i,j}A}&:=&\int_Af(\sum u_{k,l}\varphi_{k,l})
\varphi_{i,j}dx\nonumber\\
&=&\int_Af(u_{i-1,j-1}\varphi_{i-1,j-1}+
u_{i,j-1}\varphi_{i,j-1}+u_{i,j}\varphi_{i,j})\varphi_{i,j}dx.\nonumber
\end{eqnarray}
\begin{eqnarray}
I_{\Omega_{i,j}B}&:=&\int_Bf(\sum u_{k,l}\varphi_{k,l})
\varphi_{i,j}dx\nonumber\\
&=&\int_Bf(u_{i-1,j-1}\varphi_{i-1,j-1}+
u_{i-1,j}\varphi_{i-1,j}+u_{i,j}\varphi_{i,j})\varphi_{i,j}dx\nonumber
\end{eqnarray}
\begin{eqnarray}
I_{\Omega_{i,j}C}&:=&\int_Cf(\sum u_{k,l}\varphi_{k,l})
\varphi_{i,j}dx\nonumber\\
&=&\int_Cf(u_{i-1,j}\varphi_{i-1,j}+
u_{i,j}\varphi_{i,j}+u_{i,j+1}\varphi_{i,j+1})\varphi_{i,j}dx\nonumber
\end{eqnarray}
\begin{eqnarray}
I_{\Omega_{i,j}D}&:=&\int_Df(\sum u_{k,l}\varphi_{k,l})
\varphi_{i,j}dx\nonumber\\
&=&\int_Df(u_{i+1,j+1}\varphi_{i+1,j+1}+
u_{i,j}\varphi_{i,j}+u_{i,j+1}\varphi_{i,j+1})\varphi_{i,j}dx\nonumber
\end{eqnarray}
\begin{eqnarray}
I_{\Omega_{i,j}E}&:=&\int_Ef(\sum u_{k,l}\varphi_{k,l})
\varphi_{i,j}dx\nonumber\\
&=&\int_Ef(u_{i+1,j+1}\varphi_{i+1,j+1}+
u_{i,j}\varphi_{i,j}+u_{i+1,j}\varphi_{i+1,j})\varphi_{i,j}dx\nonumber
\end{eqnarray}
\begin{eqnarray}
I_{\Omega_{i,j}F}&:=&\int_Ff(\sum u_{k,l}\varphi_{k,l})
\varphi_{i,j}dx\nonumber\\
&=&\int_Ff(u_{i+1,j}\varphi_{i+1,j}+
u_{i,j}\varphi_{i,j}+u_{i,j-1}\varphi_{i,j-1})\varphi_{i,j}dx.\nonumber
\end{eqnarray}
where the sub-index $A,B,C,D,E,F$ indicate the all elements neighboring
$x_{i,j}$ in Figure 1.

Similar to the one-dimensional case, introducing the DEL 1-forms
\begin{equation}
E_{D \Omega_{i,j}}:=\{ u_{i,j+1}+u_{i+1,j}+u_{i,j-1}+u_{i-1,j}-4u_{i,j}
- I_{\Omega_{i,j}}\}du_{i,j},
\end{equation}
such that the null DEL 1-form gives rise to the equation (36) in
the finite element
method. The coboundary DEL 1-forms are given by
\begin{eqnarray}
E_{D \Omega_{i,j}} =d\a_{D \Omega_{i,j}},
\end{eqnarray}
where $ \a_{D \Omega_{i,j}}$ an arbitrary function in the function space 
on
$\Omega_{i,j}$. 
And the 
null DEL 1-form is a special case of them. 

The DEL condition now reads
\begin{equation}
dE_{D \Omega_{i,j}}=0. 
\end{equation}
Although the null DEL 1-form gives rise to the equation and satisfies 
automatically the DEL 
condition, but in general

the 
$u_{j,k}$'s are still unknown functions rather than the ones in the
solution space of the equation.

It is also straightforward to see that 

in the function space it follows from the DEL condition that
 \begin{eqnarray}
\label{fin1}
&&du_{i,j+1}\wedge du_{i,j}+du_{i+1,j}\wedge du_{i,j}
+du_{i,j-1}\wedge du_{i,j}+du_{i-1,j}\wedge du_{i,j}\nonumber\\
&=&\int_Df^\prime(u_{i+1,j+1}\varphi_{i+1,j+1}+
u_{i,j}\varphi_{i,j}+u_{i,j+1}\varphi_{i,j+1})
\varphi_{i+1,j+1}\varphi_{i,j}dx\cdot du_{i+1,j+1}\wedge du_{i,j}
\nonumber\\
&+&\int_Df^\prime(u_{i+1,j+1}\varphi_{i+1,j+1}+
u_{i,j}\varphi_{i,j}+u_{i,j+1}\varphi_{i,j+1})
\varphi_{i,j+1}\varphi_{i,j}dx\cdot du_{i,j+1}\wedge du_{i,j}\nonumber\\
&+&\int_Ef^\prime(u_{i+1,j+1}\varphi_{i+1,j+1}+
u_{i,j}\varphi_{i,j}+u_{i+1,j}\varphi_{i+1,j})
\varphi_{i+1,j+1}\varphi_{i,j}dx\cdot du_{i+1,j+1}\wedge du_{i,j}
\nonumber\\
&+&\int_Ef^\prime(u_{i+1,j+1}\varphi_{i+1,j+1}+
u_{i,j}\varphi_{i,j}+u_{i+1,j}\varphi_{i+1,j})
\varphi_{i+1,j}\varphi_{i,j}dx\cdot du_{i+1,j}\wedge du_{i,j}\nonumber\\
&+&\int_Ff^\prime(u_{i+1,j}\varphi_{i+1,j}+
u_{i,j}\varphi_{i,j}+u_{i,j-1}\varphi_{i,j-1})
\varphi_{i+1,j}\varphi_{i,j}dx\cdot du_{i+1,j}\wedge du_{i,j}\nonumber\\
&+&\int_Ff^\prime(u_{i+1,j}\varphi_{i+1,j}+
u_{i,j}\varphi_{i,j}+u_{i,j-1}\varphi_{i,j-1})
\varphi_{i,j-1}\varphi_{i,j}dx\cdot du_{i,j-1}\wedge du_{i,j}\nonumber\\
&+&\int_Af^\prime(u_{i-1,j-1}\varphi_{i-1,j-1}+
u_{i,j-1}\varphi_{i,j-1}+u_{i,j}\varphi_{i,j})
\varphi_{i,j-1}\varphi_{i,j}dx\cdot du_{i,j-1}\wedge du_{i,j}\nonumber\\
&+&\int_Af^\prime(u_{i-1,j-1}\varphi_{i-1,j-1}+
u_{i,j-1}\varphi_{i,j-1}+u_{i,j}\varphi_{i,j})
\varphi_{i-1,j-1}\varphi_{i,j}dx\cdot du_{i-1,j-1}\wedge du_{i,j}
\nonumber\\
&+&\int_Bf^\prime(u_{i-1,j-1}\varphi_{i-1,j-1}+
u_{i-1,j}\varphi_{i-1,j}+u_{i,j}\varphi_{i,j})
\varphi_{i-1,j-1}\varphi_{i,j}dx\cdot du_{i-1,j-1}\wedge du_{i,j}\nonumber\\
&+&\int_Bf^\prime(u_{i-1,j-1}\varphi_{i-1,j-1}+
u_{i-1,j}\varphi_{i-1,j}+u_{i,j}\varphi_{i,j})
\varphi_{i-1,j}\varphi_{i,j}dx\cdot du_{i-1,j}\wedge du_{i,j}\nonumber\\
&+&\int_Cf^\prime(u_{i-1,j}\varphi_{i-1,j}+
u_{i,j}\varphi_{i,j}+u_{i,j+1}\varphi_{i,j+1})
\varphi_{i-1,j}\varphi_{i,j}dx\cdot du_{i-1,j}\wedge du_{i,j}\nonumber\\
&+&\int_Cf^\prime(u_{i-1,j}\varphi_{i-1,j}+
u_{i,j}\varphi_{i,j}+u_{i,j+1}\varphi_{i,j+1})
\varphi_{i,j+1}\varphi_{i,j}dx\cdot du_{i,j+1}\wedge du_{i,j}.
\end{eqnarray}
Note that the right side is equal to the following terms that can be
denoted as $S_{A\{i,j\}},\cdots, S_{F\{i,j\}}$.
\begin{eqnarray}
S_{A\{i,j\}}
&=&(\int_Df^\prime(u_{i+1,j+1}\varphi_{i+1,j+1}+
u_{i,j}\varphi_{i,j}+u_{i,j+1}\varphi_{i,j+1})
\varphi_{i+1,j+1}\varphi_{i,j}dx+\nonumber\\
&&\int_Ef^\prime(u_{i+1,j+1}\varphi_{i+1,j+1}+
u_{i,j}\varphi_{i,j}+u_{i+1,j}\varphi_{i+1,j})
\varphi_{i+1,j+1}\varphi_{i,j}dx  
)\cdot du_{i+1,j+1}\wedge du_{i,j},\nonumber\\
S_{B\{i,j\}}
&=&(\int_Df^\prime(u_{i+1,j+1}\varphi_{i+1,j+1}+
u_{i,j}\varphi_{i,j}+u_{i,j+1}\varphi_{i,j+1})
\varphi_{i,j+1}\varphi_{i,j}dx+\nonumber\\
&&\int_Cf^\prime(u_{i-1,j}\varphi_{i-1,j}+
u_{i,j}\varphi_{i,j}+u_{i,j+1}\varphi_{i,j+1})
\varphi_{i,j+1}\varphi_{i,j}dx
)\cdot du_{i,j+1}\wedge du_{i,j},\nonumber\\  
S_{C\{i,j\}}
&=&(\int_Ef^\prime(u_{i+1,j+1}\varphi_{i+1,j+1}+
u_{i,j}\varphi_{i,j}+u_{i+1,j}\varphi_{i+1,j})
\varphi_{i+1,j}\varphi_{i,j}dx+\nonumber\\
&&\int_Ff^\prime(u_{i+1,j}\varphi_{i+1,j}+
u_{i,j}\varphi_{i,j}+u_{i,j-1}\varphi_{i,j-1})
\varphi_{i+1,j}\varphi_{i,j}dx
)\cdot du_{i+1,j}\wedge du_{i,j},\nonumber\\ 
S_{D\{i,j\}}
&=&(\int_Ff^\prime(u_{i+1,j}\varphi_{i+1,j}+
u_{i,j}\varphi_{i,j}+u_{i,j-1}\varphi_{i,j-1})
\varphi_{i,j-1}\varphi_{i,j}dx+\nonumber\\
&&\int_Af^\prime(u_{i-1,j-1}\varphi_{i-1,j-1}+
u_{i,j-1}\varphi_{i,j-1}+u_{i,j}\varphi_{i,j})
\varphi_{i,j-1}\varphi_{i,j}dx
)\cdot du_{i,j-1}\wedge du_{i,j},\nonumber\\     
S_{E\{i,j\}}
&=&(\int_Af^\prime(u_{i-1,j-1}\varphi_{i-1,j-1}+
u_{i,j-1}\varphi_{i,j-1}+u_{i,j}\varphi_{i,j})
\varphi_{i-1,j-1}\varphi_{i,j}dx+\nonumber\\
&&\int_Bf^\prime(u_{i-1,j-1}\varphi_{i-1,j-1}+
u_{i-1,j}\varphi_{i-1,j}+u_{i,j}\varphi_{i,j})
\varphi_{i-1,j-1}\varphi_{i,j}dx)\cdot du_{i-1,j-1}\wedge du_{i,j},\nonumber
\\
S_{F\{i,j\}}
&=&(\int_Bf^\prime(u_{i-1,j-1}\varphi_{i-1,j-1}+
u_{i-1,j}\varphi_{i-1,j}+u_{i,j}\varphi_{i,j})
\varphi_{i-1,j}\varphi_{i,j}dx+\nonumber\\
&&\int_Cf^\prime(u_{i-1,j}\varphi_{i-1,j}+
u_{i,j}\varphi_{i,j}+u_{i,j+1}\varphi_{i,j+1})
\varphi_{i-1,j}\varphi_{i,j}dx
)\cdot du_{i-1,j}\wedge du_{i,j}.\nonumber
\end{eqnarray}
Note that all $S_{k\{i,j\}}, k=A, \cdots, F$ are order of $h^2$.

Let us introduce two shift operators as
\begin{eqnarray}
E_1f(u_{i,j})=f(u_{i+1,j}), \nonumber\\
E_2f(u_{i,j})=f(u_{i,j+1}). \nonumber
\end{eqnarray}
Then  the following relations can be found
\begin{eqnarray}
S_{A\{i,j\}}&=&-E_1E_2S_{E\{i,j\}}, \nonumber\\
S_{B\{i,j\}}&=&-E_2S_{D\{i,j\}}, \nonumber\\
S_{C\{i,j\}}&=&-E_1S_{F\{i,j\}}, \nonumber\\
du_{i,j+1}\wedge du_{i,j}&=&-E_2(du_{i,j-1}\wedge du_{i,j}),\nonumber\\
du_{i+1,j}\wedge du_{i,j}&=&-E_1(du_{i-1,j}\wedge du_{i,j}).\nonumber
\end{eqnarray}
It is easy to check that (\ref{fin1}) becomes
\begin{eqnarray}
\label{cons}
&&(1-E_2)(du_{i,j-1}\wedge du_{i,j})+(1-E_1)(du_{i-1,j}\wedge du_{i,j})
\nonumber\\
&=&\{(1-E_2E_1)S_{E\{i,j\}}+(1-E_2)S_{B\{i,j\}}+(1-E_1)S_{F\{i,j\}} \}.
\end{eqnarray}

Let us define
\begin{eqnarray}
\omega_{D{i,j}}=du_{i-1,j}\wedge du_{i,j}-E_2 S_{E\{i,j\}}-S_{F\{i,j\}},\\
\tau_{D{i,j}}=du_{i,j-1}\wedge
du_{i,j}-S_{B\{i,j\}}-S_{E\{i,j\}}.
\end{eqnarray}
It is straightforward to show that $\omega_D$ and $\tau_D$ are
two symplectic 2-forms. Namely, they are closed with respect to $d$ on the
function space and non-degenerate. Then the
equation (\ref{cons}) is in fact a discrete version of the multisymplectic
conservation law as follows
\begin{eqnarray}
D_1 \omega_{D{i,j}} +D_2 \tau_{D{i,j}} =0. 
\end{eqnarray}
Here $D_1$ and $D_2$ are differences given by
\begin{equation}
D_1=E_1-1,\quad D_2=E_2-1, \quad
\end{equation}
and they satisfy the relation
$$
E_2E_1-1=D_1E_2+D_2.
$$

\section{Remarks}

$\quad $   In this paper we have  explored  the relations
 between symplectic and
 multisymplectic algorithms and the simple 
finite element method for the  boundary value problem of the
 semi-linear elliptic equation in one-dimensional and two-dimensional spaces. 
Although what we have found are certain simple boundary value problem of
the semi-linear  elliptic equation in lower dimensions and also quite simple 
triangulization in the
finite element method, the results still indicate that there should be very 
deep connections 
between the symplectic and/or multisymplectis algorithms and the finite
 element method. The discrete variational principle should be one of the
 key links
between these two topics. 

 We have constructed
the discrete symplectic 2-form in one-dimensional case and two discrete
symplectic 2-forms
in two-dimensional case and proved they are preserved in the sense of 
discretely
divergence free. It is important to see that these results should be able 
to generalize to higher dimensional cases and more complicated
 boundary value problem.
  The most important point in our approach is that we work with
the discrete Euler-Lagrange conditions rather than the discrete equations in the simple finite
element method. Although they are cohomologically relevant, the discrete
symplectic and discrete multisymplectic structures and their preserving
properties hold not only in the solution space but also in the function 
space in general.  

On the other hand, it should be 
mentioned that in this paper we have not employed the noncommutative
differential calculus (NCDC). But, in principle some NCDC should be 
introduced in order to establish a more complete formulation for this issue
\c{gwww00} \c{gwz00} \c{glw1}.
The reason is very simple: since  the space domain
$\Omega$ (independent variables) 
is discretized in the finite
element method, the ordinary differential calculus does not make precisely 
sense in order to construct the exterior forms in $\Omega$ and  so forth.
It should
also be mentioned that the NCDC 
to be employed is also dependant on the triangulization version in
the finite element method.

Finally, there are lots of relevant problems should to studied 
and some of them are under investigation. For instance,
what about the symplectic or multisymplectic structure-preserving properties
in the case for more complicated triangulation and boundary conditions etc.
We will publish some results on these issues elsewhere.


\vskip 15mm

\begin{thebibliography}{99}

\bibitem{fini} P. G. Ciarlet, The Finite Element Method for
Elliptic Problems, North-Holland, Amsterdam, 1978, and references therein.

Selected Works of Feng Kang (I), Ed. by Z.C. Shi et. al. (1994),
and references therein.

C. Johnson, Numerical solution of partial differential
equations by the finite element method. Cambridge
University Press, 1994, and references therein.

L.A. Ying, Notes of lectures about finite element method,
(in Chieses) Peking University, (1986).

\bibitem{symp} 

Selected Works of Feng Kang (II), Ed. by Z.C. Shi et. al. (1995),
and references therein.

J.M.Sanz-Serna and  M.P.Calvo, Numerical Hamiltonian problem,
(1994), Chapman and Hall, London, and references therein.


\bibitem{msymp}T.J. Bridges, multisymplectic structures
and wave propagations, Math. Proc. Camb. Phil. Soc., {\bf 121} (1997)147-190.

T.J. Bridges and S. Reich, multisymplectic integrators: numerical
schemes for Hamiltonian PDEs that conserve symplecticity, preprint (1999).

J.E. Marsden, G.W. Patrick and S.Shkoller, Multisymplectic geometry,
variational integrators, and nonlinear PDEs, 
Commun. Math. Phys., {\bf 199} (1998) 351-395.

\bibitem{glw1}H.Y. Guo Y.Q. Li and K. Wu, On symplectic and multisymplectic 
structures and their discrete versions in Lagrangian formalism, 
ITP-preprint (March, 2001) hep-ph/0104064.

\bibitem{glw2}H.Y. Guo Y.Q. Li and K. Wu, A note on symplectic
algorithms, ITP-preprint (March, 2001) physics/0104030.

\bibitem{vese} A.P. Veselov, Integranble discrete-time
system and difference operator,
Funkts. Anal. Prilozhen., {\bf 22} (1988)1-13.

J. Moser and A.P. Veselov, Discrete versions of some classical
integrable systems and factorization of matrix polynomials,
Commun. Math. Phys.,
{\bf 139} (1991)217-243.

\bibitem{mars}
J.M. Wendlandt and J.E. Marsden, Mechanical integrators derived
from a discrete variation plinciple, Physica {\bf D 106} (1997)223-246.

\bibitem{qin} Y.J. Sun and M.Z. Qin, Variational integrators
and application for higher order differential equations,
CCAST-WL workshop series: {\bf 118}, 45-58.


\bibitem{gwww00} H.Y. Guo, K.Wu, S.H. Wang, S.K. Wang and G.M.Wei,
 Noncommutative Differential
Calculas Approach to Symplectic Algorithm on Regular Lattice, 
Comm. Theor. Phys., {\bf 34} (2000) 307-318.


\bibitem{gwz00} H.Y.Guo, K.Wu and W.Zhang, Noncommutative Differential 
Calculus on
Abelian Groups and Its Applications, Comm.Theor. Phys., {\bf 34} (2000) 245-250.

\bibitem{ji} X.M. Ji, to appear.

\end{thebibliography}
\end{document}